\documentclass[twocolumn]{article}

\usepackage{eufrak,dsfont}

\begin{document}

\title{Intrinsic Time Deparameterization of The Canonical Connection Dynamics of General Relativity}          % Enter your title between curly braces
\author{Vasudev Shyam\\ \\Centre for Fundamental Research and Creative Education,\\
Bangalore, India}        % Enter your name between curly braces
\date{\today}          % Enter your date or \today between curly braces
\maketitle
\begin{abstract}
We investigate the implications of intrinsic time deparameterization (first dealt with in \cite{hco1}, \cite{hc1}) on the phase space of the connection representation of canonical gravity in the form of the Ashtekar variables. We find that, much like the metric representation of this formalism, the Hamiltonian constraint now becomes a physical Hamiltonian generating evolution with respect to an intrinsic time. The complete observables for the theory are then constructed. Also, the dynamics in this formulation is cast into a classical master constraint theory.\end{abstract}
\section{Introduction}       % Enter section title between curly braces
In this paper we shall arrive at the Ashtekar variables for the system described by the variables
$$\bar{\gamma}_{ab}=\gamma^{-1/3}\gamma_{ab},$$
which is unimodular,
and
$$\sigma^{ab}=\gamma^{1/3}(\pi^{ab}-\frac{1}{3}\gamma^{ab}\textrm{tr}\pi).$$
This is nothing but a conformal decomposition of the canonical variables of Hamiltonian General Relativity where the conformal factor is  specialized to $\phi^{4}=\gamma^{-1/3}$. We shall proceed with this form of the canonical variables despite the conformal factor being a density. This choice of variables facilitates the following split of the symplectic potential
\begin{equation}\int \textrm{d}^{3}x \pi^{ab}\textrm{d}_{\delta}\gamma_{ab}=\int \textrm{d}^{3}x \sigma^{ab}\textrm{d}_{\delta}\bar{\gamma}_{ab}+\textrm{tr}\pi \textrm{d}_{\delta}\textrm{ln}\gamma^{1/3}.\end{equation}
With this, it is possible to deparameterize canonical general relativity. In this paper, we attempt to do the same for the Ashtekar variables. In order to do so, we shall begin not with the (conformally decomposed) canonical ADM variables described above, but with the following variables instead
\begin{equation}\bar{\gamma}_{ab}=\gamma^{-1/3}\gamma_{ab}\end{equation}
\begin{equation}\textrm{tr}K=\gamma^{ab}K_{ab}\end{equation}
\begin{equation}\bar{K}_{ab}=\gamma^{-1/3}\left(K_{ab}-\frac{1}{3}\textrm{tr}K\right)\end{equation}
\subsection{The Ashtekar Variables}    % Enter subsection title between curly braces
In this section the Ashtekar variables in the conformal decomposition described above, for the construction in the case of a general conformal factor see \cite{chtw1}, \cite{chtw2} (and references therein). We begin with the triad decomposition
$$e_{ai}=\textrm{det}(e)^{1/3}\bar{e}_{ai},$$
where
$$\textrm{det}(e)=\textrm{det}e_{ai}$$
Thus the densitized triad in the Ashtekar variables is given as
$$E^{a}_{i}=\textrm{det}(e)^{2/3}\bar{E}^{a}_{i}$$
The extrinsic curvature part of the Ashtekar connection is given by
$$K^{i}_{a}=\textrm{det}(e)^{-2/3}\bar{K}^{i}_{a}+\frac{1}{3}\textrm{det}(e)^{1/3}\textrm{tr}K\bar{E}^{i}_{a}$$
$$=>\bar{K}^{i}_{a}=\textrm{det}(e)^{2/3}\left(K^{i}_{a}-\frac{1}{3}\textrm{det}(e)^{1/3}\textrm{tr}K\bar{E}^{i}_{a}\right)$$
The Ashtekar connection is thus given by
\begin{equation}\bar{A}^{i}_{a}=\bar{\Gamma}^{i}_{a}+\bar{K}^{i}_{a}.\end{equation}
It is important to note that this could also be written as
$$\bar{A}^{i}_{a}=\bar{\Gamma}^{i}_{a}+\textrm{det}(e)^{2/3}\left(K^{i}_{a}-\frac{1}{3}\textrm{det}(e)^{1/3}\textrm{tr}K\bar{E}^{i}_{a}\right),$$
notice how the conformal factor appears where the Barbero-Immirzi parameter usually would. 
\subsection{Constraints and the Hamiltonian}
The Gauss constraint is given by
\begin{equation}\mathcal{G}_{i}=\bar{D}_{a}\bar{E}^{a}_{i}=0.\end{equation}
The diffeomorphism constraint now looks like
\begin{equation}\mathcal{D}_{a}=\frac{4}{3}\textrm{det}(e)^{-1}\textrm{tr}K,_{a}+\bar{F}^{j}_{ab}\bar{E}^{b}_{j}-\bar{A}^{j}_{a}\mathcal{G}_{j}=0.\end{equation}
Here $\bar{F}=d\bar{A}+\bar{A}\wedge \bar{A}$ is the curvature of the Ashtekar connection.
The Hamiltonian constraint now takes the following form
\begin{equation}H=-\frac{1}{3}\textrm{tr}K^{2}+\textrm{det}(e)^{-2}[e^{4/3}\bar{F}^{k}_{ab}\epsilon^{ijk}$$\\$$-\frac{4\textrm{det}(e)^{4/3}+1}{2}\epsilon^{abc}\bar{K}^{i}_{b}\bar{K}^{j}_{c}]\bar{E}^{a}_{i}\bar{E}^{b}_{j}+$$\\$$\textrm{det}(e)^{-5/6}\bar{\Delta}\textrm{det}(e)^{1/6}=0\end{equation}
It is easy to see that this can now be written as
\begin{equation}\frac{1}{\sqrt{3}}\textrm{tr}\tilde{K}=$$\\$$ \sqrt{\left[\textrm{det}(e)^{4/3}\bar{F}^{k}_{ab}\epsilon^{ijk}-\frac{4\textrm{det}(e)^{4/3}+1}{2}\epsilon^{abc}\bar{K}^{i}_{b}\bar{K}^{j}_{c}\right]\bar{E}^{a}_{i}\bar{E}^{b}_{j}+R_{\phi}}$$\\$$=-\mathfrak{h}.\end{equation}
Here $$\textrm{tr}\tilde{K}=\textrm{det}(e)\textrm{tr}K,$$ and $$R_{\phi}=8\textrm{det}(e)^{-1/2}\bar{\Delta}\textrm{det}(e)^{1/6}.$$
The latter term reflects the locality of the conformal factor.
This is an effective Hamiltonian and not a constraint. It is a remarkable coincidence that this form of the true Hamiltonian in this `internally' unimodular gauge of conformal gravity is the very similar to what has been derived in the geometrodynamic theory which is given by
\begin{equation}-\frac{\textrm{tr}\pi}{\sqrt{6}}=\sqrt{\bar{G}_{abcd}\sigma^{ab}\sigma^{cd}-\gamma R}.\end{equation} 
\section{The Intrinsic Time Deparameterization}
In this section the deparameterization of the phase space described by the above Hamiltonian system is attempted. First notice that the L.H.S of the above Hamiltonian, i.e. $\textrm{tr}\tilde{K}$ is canonically conjugate to the variable $\textrm{ln}[\textrm{det}(e_{ai})]^{2/3}$. Thus we can write an additional symplectic potential as
$$\int \textrm{d}^{3}x \textrm{tr}\tilde{K} \textrm{d}_{\delta}\tau.$$
Here $\tau=\textrm{ln}[\textrm{det}(e_{ai})]^{1/3}$. this shall serve as the intrinsic time. Now, the total presymplectic potential is given by
\begin{equation}\Theta=\int \textrm{d}^{3}x \bar{E}^{a}_{i}\textrm{d}_{\delta}\bar{A}^{i}_{a}+\textrm{tr}\tilde{K} \textrm{d}_{\delta}\tau \end{equation}
$$=>\Theta=\int\textrm{d}^{3}x \bar{E}^{a}_{i}\textrm{d}_{\delta}\bar{A}^{i}_{a}-\mathfrak{h} \textrm{d}_{\delta}\tau  $$
The constraint  surface $\tilde{\Gamma}$ of the phase space $\Gamma$ is defined by     
$$\mathcal{G}_{i}[\Lambda]=\int_{\Sigma}\mathcal{G}^{i}\Lambda_{i}=0,$$
and
$$\mathcal{D}_{a}[\xi^{a}]=\int_{\Sigma}\xi^{a}\mathcal{D}_{a}=0.$$
And so, we have the presymplectic equation
\begin{equation}\iota_{\mathcal{X}}\textrm{d}_{\delta}\Theta|_{\tilde{\Gamma}}=0.\end{equation}
The Hamiltonian vector field that satisfies this is given by
$$\mathcal{X}=\frac{\delta}{\delta \tau}-\frac{\sqrt{3}}{2\textrm{tr}\tilde{K}}((\textrm{det}(e)^{4/3}\epsilon_{ijk}\bar{F}^{j}_{ab}\bar{E}^{b}_{k}+$$\\$$\frac{4\textrm{det}(e)^{4/3}+1}{2}\bar{K}^{i}_{[a}\bar{K}^{j}_{b]}\bar{E}^{j}_{b}$$\\$$-\bar{D}_{a}\Lambda+\mathcal{L}_{\xi^{a}}\bar{A}^{i}_{a}+\bar{D}_{a}\bar{A}(\xi))\frac{\delta}{\delta \bar{A}^{i}_{a}}$$\\$$+(\textrm{det}(e)^{4/3}\bar{D}_{a}\epsilon^{ijk}\bar{E}^{a}_{i}\bar{E}^{b}_{j}+$$\\ $$\frac{4\textrm{det}(e)^{4/3}+1}{2}\epsilon^{abc}\bar{K}^{i}_{a}\bar{E}^{a}_{i}\bar{E}^{b}_{j}+$$\\$$\epsilon^{ijk}\Lambda_{j}\bar{E}^{b}_{k}+\mathcal{L}_{\xi^{a}}\bar{E}^{a}_{i}+ \epsilon^{ijk}\Lambda_{j}(\bar{A}(\xi))_{k})\frac{\delta}{\delta \bar{E}^{a}_{i}}).$$
Here $\bar{A}(\xi)=\bar{A}_{a}\xi^{a}.$
This is a suspended Hamiltonian vector field and it has the property
$$\mathcal{X}(\tau)=1.$$
Now, we see that this vector field, on acting on the deparameterized phase space co ordinates
\[z^{I}=\left(\begin{array}{cc}\bar{E}^{a}_{i}\\ \bar{A}^{i}_{a}\end{array}\right),\]
gives
\[\frac{\delta}{\delta \tau}z^{I}=\left(\begin{array}{cc}-\frac{\delta \mathfrak{h}[\bar{A},\bar{E}]}{\delta \bar{A}^{i}_{a}}\\ \frac{\delta \mathfrak{h}[\bar{A},\bar{E}]}{\delta \bar{E}^{a}_{i}}\end{array}\right).\]
This implies that the (locally) Hamiltonian vector field $\mathcal{X}$ can be written as 
$$\mathcal{X}=\frac{\delta}{\delta \tau}-\mathcal{X}_{\mathfrak{h}}.$$
The presymplectic equation, on the deparameterized phase space, now becomes
\begin{equation}\iota_{\mathcal{X}_{\mathfrak{h}}}\textrm{d}_{\delta}(\bar{E}^{a}_{i}\textrm{d}_{\delta}\bar{A}^{i}_{a})=\textrm{d}_{\delta}\mathfrak{h}.\end{equation}
Assuming initially that the phase space is given by the double $(\tilde{\Gamma},\tilde{\Omega}:=\textrm{d}_{\delta}\Theta|_{\tilde{\Gamma}}),$ then, we can reduce this phase space by choosing a `time function' $T:\tilde{\Gamma}\rightarrow \mathds{R}.$ With this we can view $\tilde{\Gamma}$ as a product of and interval $I \subset \mathds{R}$ which is the range of $T$ with the deparameterized phase space $\bar{\Gamma}.$ We have the map $\varphi$ with
$$\varphi:I \times \bar{\Gamma}\rightarrow \tilde{\Gamma}.$$ 
this satisfies 
$$T(\varphi(\tau,f(z^{I}))=\tau,\tau\in I.$$
We can now employ this map to pull back the presymplectic form to the deparameterized phase space, i.e.
$$\varphi^{*}\Theta=\Theta_{\tau}-\mathfrak{h}\textrm{d}_{\delta}\tau,$$
here $\Theta_{\tau}$ is the symplectic potential on $\bar{\Gamma}.$
Surely enough
$$\iota_{\frac{\delta}{\delta \tau}}\varphi^{*}\Theta=-\mathfrak{h},$$
holds. Note that $\textrm{d}_{\delta}\Theta_{\tau}$ is strongly non degenerate, thus it satisfies
$$\iota_{\mathcal{X}_{\mathfrak{h}}}\textrm{d}_{\delta}\Theta_{\tau}=\textrm{d}_{\delta}\mathfrak{h}.$$
This is precisely what eq(13) tells us. Thus $(I\times \bar{\Gamma},\Theta_{\tau},\mathfrak{h})$ is the deparameterization of $(\tilde{\Gamma},\tilde{\Omega}).$
\section{Complete Observables}
We could have alternatively written the new form of the Scalar Hamiltonian constraint as an Abelian first class constraint
$$\mathfrak{C}=\frac{1}{\sqrt{3}}\textrm{tr}\tilde{K}-\mathfrak{h}.$$
and on its constraint surface we can define the constrained, but not totally constrained Hamiltonian
$$\bar{\mathcal{H}}=\int_{\Sigma} \mathfrak{h}+\xi^{a}\mathcal{D}_{a}+\Lambda_{i}\mathcal{G}^{i}$$
This is Abelian since it is linear in the momenta and the true Hamiltonian also Poisson-commutes with itself. The flow of some phase space function $f[z^{I}]$ is a function that satisfies 
$$\frac{\textrm{d}}{\textrm{d}\tau}\alpha^{\tau}_{\bar{\mathcal{H}}}(f)[z^{I}]=\mathcal{X}_{\bar{\mathcal{H}}}\alpha^{\tau}_{\bar{\mathcal{H}}}(f)[z^{I}],$$
and
$$\alpha^{0}_{\bar{\mathcal{H}}}(f)[z^{I}]=f[z^{I}],$$
Also
$$\alpha^{\tau}_{\bar{\mathcal{H}}}(f)[z^{I}]=f[\alpha^{\tau}_{\bar{\mathcal{H}}}(z^{I})].$$
Thus, we see that a flow can be written explicitly as
$$\alpha^{\tau}_{\bar{\mathcal{H}}}(f)[z^{I}]=\sum_{n=0}^{\infty}\frac{\tau^{n}}{n!}\mathcal{X}^{n}_{\bar{\mathcal{H}}}(f).$$
In general one can attain a complete observable as follows 
$$F^{\tau}_{f,T}(z^{I}):=\alpha^{\tau}_{C}[z^{I}]|_{\alpha^{\tau}_{C}[T](z)=\tau},$$
where $C$ is a constraint such that if it Poisson commutes with any phase space function, that function is known as a Dirac  Observable. $T$ is known as a clock variable which fixes its value $T=\tau$. In the case of the system considered here, the complete observables are given by
$$F^{\tau}_{f,\scriptsize\textrm{ln(det}(e))^{2/3}}=\sum^{\infty}_{n=0}\frac{(\tau-\textrm{ln}(\textrm{det}(e))^{2/3})^{n}}{n!}\mathcal{X}^{n}_{\bar{\mathcal{H}}}$$
It's evolution is governed by the equation
$$\frac{\partial}{\partial \tau}F^{\tau}_{f,\scriptsize\textrm{ln(det}(e))^{2/3}}=\mathcal{X}_{\mathfrak{h}}F^{\tau}_{f,\scriptsize\textrm{ln(det}(e))^{2/3}.}$$
\section{The Classical Master Constraint}
The Classical Master constraint for this theory can be written as 
$$\mathbf{M}=\frac{1}{2}\int_{\Sigma} \frac{\mathfrak{C}^{2}}{\sqrt{\gamma}}=\frac{1}{2}\int_{\Sigma}\frac{(\frac{1}{\sqrt{3}}\textrm{tr}\tilde{K}-\mathfrak{h})^{2}}{\sqrt{\gamma}}.$$
The constraint surface associated to it is given by
$$\Gamma_{\mathbf{M}}:=\left\{z\in \Gamma|\mathbf{M}(z)=0\right\}$$
We then have the following first class constraint algebra on $\Gamma_{\mathbf{M}}$
$$\left\{\mathbf{M},\mathbf{M}\right\}=0$$
$$\left\{\mathcal{D}_{a}(\xi^{a}),\mathbf{M}\right\}=0$$
$$\left\{\mathcal{G}^{i}(\Lambda_{i}),\mathbf{M}\right\}=0.$$
A weak Dirac observable $O$ is defined to be a phase space function that satisfies
$$\left\{O,\left\{O,\mathbf{M}\right\}\right\}|_{\Gamma_{\mathbf{M}}}=0.$$
This constraint has, associated with it, a Hamiltonian vector field given by
$$(\mathcal{X}_{\mathbf{M}})^{\flat}|_{\Gamma_{\mathbf{M}}}=\textrm{d}_{\delta}\mathbf{M}.$$
With this we can find a flow
$$\alpha^{\tau}_{\mathbf{M}}=\sum^{n=\infty}_{n=0}\frac{\tau^{n}}{n!}\mathcal{X}^{n}_{\mathbf{M}}.$$
A strong Dirac observable is now defined as one which satisfies
$$\alpha^{\tau}_{\mathbf{M}}[O]=O,\forall \tau\in \mathds{R}$$
If one wished to treat even the diffeomorphism constraint on an equal footing, one could choose to deal with the extended master constraint, defined as
$$\mathbf{M}_{E}=\int_{\Sigma}\frac{(\frac{1}{\sqrt{3}}\textrm{tr}K-\mathfrak{h})^{2}+\gamma^{ab}\mathcal{D}_{a}\mathcal{D}_{b}}{\sqrt{\gamma}}.$$
One can also choose to treat all the constraints on the same footing, in which case we have the total master constraint
$$\mathbf{M}_{T}=\int_{\Sigma}\frac{(\frac{1}{\sqrt{3}}\textrm{tr}K-\mathfrak{h})^{2}+\gamma^{ab}\mathcal{D}_{a}\mathcal{D}_{b}+\mathcal{G}_{i}\mathcal{G}^{i}}{\sqrt{\gamma}}.$$
With respect to the above, one may also define a master action given by
$$S_{\mathbf{M}}=\int_{\Sigma}\textrm{d}\tau\left(\left\{ \bar{E}^{a}_{i}\dot{\bar{A}}^{i}_{a}\right\}-\mathbf{M}_{T}\right) $$
Note that in this formalism, there are no constraints anymore.
\section{Conclusion}
Thus we have presented the classical framework of intrinsic time gravity in the connection representation. The general case, for any conformal factor is explored in \cite{chtw1}, but the linchpin for the deparameterization lies in the $\gamma^{-1/3}$ choice of the conformal factor, and through it we attain the intrinsic time, and this was first noticed in \cite{hco1} and \cite{hc1}. The Master constraint formalism that has been presented here shall be especially useful for the quantization of this theory. Also, the master action can be used to construct a spinfoam model which has a sound relationship with the canonical framework, although it can't possibly be covariant.
\section{Acknowledgments}
This work was carried out at the \textit{Centre for Fundamental Research and Creative Education}, Bangalore, India. The author would like to acknowledge his mentor Dr. B.S Ramachandra for invaluable guidance and the Director Ms. Pratiti B R for facilitating an atmosphere of free scientific inquiry so conducive to creativity. The author also wishes to acknowledge fellow researchers Magnona H Shastry, Madhavan Venkatesh, Karthik T Vasu and Arvind Dudi.

% Set the ending of a LaTeX document

\begin{thebibliography}{100}
\bibitem{hco1} Niall Ó Murchadha, Chopin Soo, Hoi-Lai Yu, Intrinsic time gravity and the Lichnerowicz-York equation,  arXiv:1208.2525v1
\bibitem{hc1} Chopin Soo, Hoi-Lai Yu General Relativity without paradigm of space-time covariance: sensible quantum gravity and resolution of the problem of time,  arXiv:1201.3164v2
\bibitem{chtw1} Charles H.-T. Wang, Towards conformal loop quantum gravity, aXiv:gr-qc/0512023
\bibitem{chtw2} Charles H.-T. Wang, Unambiguous spin-gauge formulation of canonical general relativity with conformorphism invariance, arXiv:gr-qc/0507044
\bibitem{tt1} Thomas Thiemann, The Phoenix Project: Master Constraint Programme for Loop Quantum Gravity, arXiv:gr-qc/0305080
\bibitem{jt1} Johannes Tambornino, Relational Observables in Gravity: a Review, arXiv:1109.0740 [gr-qc]
\bibitem{ehl}R. Beig in Canonical Gravity: From Classical To Quantum, Springer Verlag, J.Ehlers, H.Friedrich: The classical theory of canonical general relativity
\end{thebibliography}
\end{document}